# Recrystallization as a tool to characterize and control the properties of cold sprayed materials: case study on 316L steel

**Petr Haušild[1], Jaroslav Čech[1], Petr Králíček[2,3], Tomáš Heller[1,2], Jan Cizek[2,3]**

[1]Czech Technical University in Prague, Faculty of Nuclear Sciences and Physical Engineering, Department of Materials, Trojanova 13, 120 00 Prague 2, Czech Republic

[2]Institute of Plasma Physics of the Czech Academy of Sciences, Department of Materials Engineering, Za Slovankou 1782/3, 182 00 Prague 8, Czech Republic

[3]Czech Technical University in Prague, Faculty of Mechanical Engineering, Department of Materials Engineering, Karlovo Namesti 13, 120 00 Prague 2, Czech Republic

**Abstract**

Owing to the unique buildup mechanisms, recrystallization in cold sprayed materials significantly differs from that in conventional counterparts. In this study, the relationship between deformation-induced microstructural heterogeneity, recrystallization behavior, and mechanical properties of cold sprayed materials was investigated using EBSD and nanoindentation. The 316L stainless steel was selected as a model material. The deposit was characterized in its as-sprayed condition and after annealing at temperatures between 900 and 1100 K. Thermomechanical estimates suggest that locally, temperature can rise over 1000 K and very high dislocation densities (up to $10^{17}$ $m^{-2}$) form at particle boundaries during impact, providing a strong driving force for subsequent recrystallization. The EBSD observations confirmed the intense defect accumulation and grain fragmentation localized near the particle interfaces. Recrystallization initiated preferentially in these highly strained areas after annealing at 1000 K and progressed with increasing temperature and holding time, leading to complete recrystallization after 1 h at 1100 K. The recrystallized grain-size distribution was well described by a heavy-tailed log-normal distribution, which indicates extremely heterogeneous strain partitioning generated during particle impacts. Nanoindentation measurements showed that the high hardness of the cold sprayed deposit remained largely unchanged after annealing at 900 K but gradually decreased as recrystallization progressed. We further demonstrate that controlled post-spray recrystallization can serve as a powerful tool for characterizing deformation during deposition and for tailoring the microstructure and mechanical properties of cold-sprayed materials.

**Novelty Statement**

This work demonstrates that recrystallization can be used as a powerful tool for unraveling the highly heterogeneous deformation introduced during cold spray deposition. By combining EBSD and nanoindentation, the study establishes a link between localized strain accumulation, deformation history, and mechanical property evolution in cold-sprayed 316L stainless steel through recrystallization kinetics and post-annealing grain size distribution analysis.

**Introduction**

Cold spray (CS), also known as kinetic spraying, is now considered an established method for deposition of metallic materials (Ref 1-4). Unlike in conventional thermal spraying methods, the starting material in CS does not heat to high temperatures during the spray process, and the particles remain in a solid state when they impact the substrate. CS reduces oxidation, phase and compositional changes, and other typical problems associated with thermal spraying. Thanks to these advantages, CS can be used for both coating and repairs, as well as in additive manufacturing, a technology that is currently in high demand.

Alongside shock wave propagation (Ref 5), oxide layer break-up and jetting (the intense outward ejection of material at the interface between particle and substrate), adiabatic shear instability (ASI) has emerged as one of the central concepts in understanding bonding mechanisms in CS (Ref 6). ASI is a deformation phenomenon that occurs at very high strain rates, typically on the order of $10^6$–$10^9$ $s^{-1}$ (Ref 6, 7), where the heat generated by plastic work cannot be dissipated fast enough and remains locally confined. Under such conditions, thermal softening can balance strain and strain-rate hardening, leading to the formation of narrow zones of intense shear deformation. In classical applications, such as ballistic impact or machining, ASI is often associated with failure in the form of shear bands. However, in interfacial high-velocity impact processes, such as explosive welding and magnetic pulse welding (Ref 8), this same instability promotes bonding rather than fracture by enabling material flow via localized plastic deformation and thermal softening.

In these processes, ASI-driven deformation results in the formation of a high-velocity material jet that removes oxide layers and contaminants from the interface. A similar mechanism is proposed for cold spray: the localized shear instability induces severe plastic flow and jetting, exposing fresh metallic surfaces that come into intimate atomic contact (Ref 9). This process allows metallurgical bonding to occur in the solid state, without the need for melting or diffusion over long timescales. Although local transient melting may occur under extreme conditions (Ref 10), it is generally not considered a prerequisite for bonding.

An essential characteristic of ASI in cold spray is its highly localized nature. Both modelling and microstructural analyses show that the instability initiates at the outer rim of the particle-substrate or particle-particle contact zone, where shear stresses are highest, and propagates inward as the impact velocity increases (see, e.g., (Ref 11)). Importantly, only certain regions of the incoming particle undergo this instability, while their bottom regions—characterized by high hydrostatic pressure but low shear—often remain unbonded, giving rise to the so-called south pole phenomenon (Ref 12). This spatial heterogeneity explains several experimentally observed features, including jet-like material extrusion at the periphery of impact sites, vortex-like shear mixing (Ref 13), partial bonding across interfaces, and the relatively lower cohesive strength of cold-sprayed deposits compared to bulk materials.

The occurrence of ASI is governed by a combination of material properties and process parameters. The key material factors include flow stress, strain-rate sensitivity, thermal softening behavior, and melting temperature, while the process parameters primarily include particle size and morphology and in-flight particle velocity and temperature. Each of these factors strongly influences the thermomechanical response during impact. For instance, higher impact velocities and temperatures promote ASI by increasing the rate of plastic dissipation and the magnitude of adiabatic heating (Ref 11). Conversely, smaller particles or systems with efficient heat conduction tend to suppress instability due to reduced adiabaticity.

Despite the substantial progress in understanding ASI in cold spray, several open questions remain. The exact relationship between the extent of the shear-unstable region and the resulting bond strength is not entirely resolved, and the role of possible interfacial phase transformations or localized melting is still debated (Ref 10). Additionally, the applicability of ASI-based bonding mechanisms to non-ductile materials, such as ceramics or metallic glasses, is less clear, as these systems may exhibit alternative deformation and bonding mechanisms. Limitations in constitutive models at ultra-high strain rates further complicate quantitative predictions (Ref 14).

The final microstructure of CS deposits is governed by a combined effect of large shear strains, high strain rates, and rising temperature. Strain localization in a narrow interfacial region leads to grain subdivision due to dislocation accumulation. Very high local strains (often exceeding 1000%), extreme strain rates, and temperature can trigger dynamic or post-dynamic recrystallization when the thermal softening exceeds strain hardening (Ref 11). All these events result in the formation of ultrafine equiaxed grains or nanograins near the interfaces. An indirect connection between adiabatic shear instability and dynamic recrystallization in cold spray coatings can be found in the literature (Ref 11, 15-17). However, the exact mechanisms (true dynamic recrystallization vs. recovery vs. melting-assisted processes) remain an open research question.

In this paper, we present several novel insights into the recrystallization occurring in CS deposits based on results obtained by a combination of electron backscatter diffraction (EBSD) and nanoindentation methods. After characterizing the model 316L steel in the as-sprayed condition, annealing was performed to link the evolution of the microstructure and hardness with the distribution of strain in individual particles. Based on the size distribution of the recrystallized grains, the conditions that could lead to the formation of this characteristic microstructure are inferred.

## Experimental setup

Owing to its suitable mechanical, physical, and chemical properties, 316L austenitic stainless steel was chosen as a model material. The steel was selected due to its high chemical stability that prevents post-spraying oxidation and its low to intermediate value of stacking fault energy of ~30 $mJ/m^2$ (Ref 18, 19) which promotes grain refinement during large strain deformation and inhibits the martensitic transformation (compared to, e.g., 304L austenitic stainless steels) (Ref 20). Over a broad temperature range, the material is also less strain-rate sensitive than other metals such as titanium alloys or superalloys, which makes estimations less inaccurate in the high strain-rate regime where the flow stress relationships are unknown.

The feedstock powder was a commercially available 316L powder (Sandvik Osprey, UK) manufactured by inert gas atomization. According to the manufacturer, the particle size range was 15–45 µm. The nominal chemical composition of the powder is given in Table 1.

Table 1. Nominal chemical composition of the powder (wt.%).

| Fe | C | Cr | Ni | Mo | Si | Mn | S | P | N |
|---|---|---|---|---|---|---|---|---|---|
| balance | ≤ 0.03 | 16.0-18.0 | 10.0-14.0 | 2.0-3.0 | ≤ 1.0 | ≤ 2.0 | ≤ 0.03 | ≤ 0.045 | ≤ 0.10 |

The particle size distribution of the feedstock powder was analyzed using the wet laser diffraction method (Mastersizer 3000, Malvern Panalytical, UK). The powder was deposited using a high-pressure Kinetiks 8000-X system (CGT, Germany) with nitrogen as the process gas. The used gas pressure and temperature were 3.5 MPa and 1000 °C, respectively. A 160-mm convergent-divergent SiC OUT1 nozzle with an expansion ratio of 5.6 was used. A robot traverse speed of 300 mm/s was used, and the stand-off distance was maintained at 30 mm during the deposit thickness growth. At this distance, the average in-flight velocity of the powder particles was 650 m/s as measured by the Accuraspray CS sensor (Tecnar, Canada).

The samples were characterized in their as-sprayed condition and after annealing in an electric furnace in air followed by oil quenching. The annealing times and temperatures were chosen as 1 h at 900 K (~630 °C), 1 h at 1000 K (~730 °C), 4 h at 1000 K, and 1 h at 1100 K (~830 °C). As an aluminum substrate was used, the deposits were detached after spraying to prevent contamination due to the substrate melting.

The samples were prepared by standard procedure (cutting, mechanical grinding, and polishing) with the final electro-polishing step (using 5% perchloric acid and a voltage of 40 V) to obtain a stress-free surface layer for subsequent EBSD and nanoindentation analyses.

EBSD analysis was carried out in an IT500HR (Jeol, Japan) scanning electron microscope (SEM) equipped with an EDAX Velocity camera (Ametek, US). The electropolished samples were inspected at a tilt angle of 70°. The data acquired at a step size of 100 nm were evaluated by EDAX OIM Analysis™ software using spherical indexing. The results are presented in the form of Correlation Index (CI, a measure of how well an acquired Kikuchi diffraction pattern matches a simulated pattern for a given crystal orientation, useful for reliable grain/subgrain boundary detection), Inverse Pole Figure (IPF, showing crystallographic directions aligned with normal sample direction), and Kernel Average Misorientation maps (KAM, the average misorientation angle between a measurement point and its neighboring points which can be converted to the density of geometrically necessary dislocations using Nye's dislocation theory (Ref 21)). The grain orientation spread (GOS) parameter, i.e., internal misorientation within an individual grain, was used to quantify the fraction of recrystallized, strain-free grains.

The indentation measurements were performed at room temperature using the Anton Paar $NHT^2$ nanoindentation tester (Anton Paar TriTec, Switzerland) equipped with a Berkovich indenter using the instrumented indentation technique (Ref 22) based on the Oliver-Pharr method (Ref 23).

Here, the contact depth ($h_c$) is obtained from the maximum load ($P_{max}$) and depth ($h_{max}$) as:

$$h_c = h_{max} - \varepsilon \frac{P_{max}}{\frac{dP}{dh}}, \quad (1)$$

where $\varepsilon$ is a geometric constant equal to ~0.75 (this value was refined by a methodology based on the unloading power-law exponent presented in (Ref 24)).

The tip area function $A(h_c)$ of the Berkovich indenter was calibrated using the standard procedure and a fused silica etalon in the following form with fitting parameters $C_1, \ldots, C_5$:

$$A(h_c) = 24.5h_c^2 + C_1 h_c^1 + C_2 h_c^{1/2} + C_3 h_c^{1/4} + C_4 h_c^{1/8} + C_5 h_c^{1/16}. \quad (2)$$

The hardness ($H$) is then given as:

$$H = \frac{P_{max}}{A(h_c)}. \quad (3)$$

To obtain hardness at different indentation depths, so-called continuous multi-cycle (CMC) indentation consisting of continuously increasing loading and partial unloading was performed. At each indentation, 20 such cycles were carried out, with the maximum load of 1 mN in the first cycle and 400 mN in the last cycle. The increases in the maximum load between the cycles were selected to create an approximately equidistant distribution of the indentation depths $h_c$. Each cycle consisted of 10 s loading time, followed by a 5 s hold at the respective maximum load, and unloading to 10% of the maximum cycle load in 10 s. At least seven CMC indentations were performed for each material condition.

From the Sneddon solution (Ref 25), the contact stiffness is related to reduced modulus and contact area (radius) as:

$$\frac{dP}{dh} = \frac{2}{\sqrt{\pi}} E_r \beta \sqrt{A(h_c)}, \quad (4)$$

where $\beta$ is a constant dependent on the indenter geometry (~1.034 for a three-sided pyramid) and the reduced modulus ($E_r$) comes from the Young's moduli of the sample and the indenter, $E$ and $E_i$, and the respective Poisson ratios $\nu$ and $\nu_i$:

$$\frac{1}{E_r} = \frac{1-\nu^2}{E} + \frac{1-\nu_i^2}{E_i}. \quad (5)$$

The uneven localization of deformation into the small regions requires local measurements, i.e., at shallow indentation depths. However, at shallow depths, the indentation size effect, i.e., size-dependent increase in hardness must be taken into account, which makes it difficult to compare mechanical behavior at this scale. For such comparison, a single value of hardness is not sufficient—the entire hardness depth profile should be compared. The CMC method allows only a limited number of measurements at shallow depths due to the necessary intervals between individual cycles, and its use could potentially affect the dislocation structure through cyclic loading and unloading, particularly when an excessive number of cycles is employed. To assess the hardness depth profile locally in very small regions (e.g., severely deformed) and without the cyclic loading-unloading, we also adopted a second methodology based on (single) constant strain rate indentation tests ($\dot{h}/h = 0.01\ \mathrm{s}^{-1}$) and the assumption of constant and known reduced modulus $E_r$ tabulated or obtained from, e.g., the final unloading step and Equation 4. Combining equations 1, 2, 4, and 5 to eliminate the contact stiffness, the contact depth can then be (iteratively) computed for each depth by solving the non-linear equation:

$$h_c(h) = h - \varepsilon \frac{\sqrt{\pi} P}{2 E_r \beta \sqrt{A(h_c)}}. \tag{6}$$

Finally, a rectangular test specimen of dimensions 3×3×2 $\mathrm{mm}^3$ was cut from the fully recrystallized sample (annealed at 1100 K for 1 h) and subjected to a compression test at a strain rate of ~$10^{-4}\ \mathrm{s}^{-1}$ using an Inspekt 100 kN testing machine (Hegewald & Peschke, Germany).

## Results and discussion

The laser diffraction measurement determined both the volume and number size distributions of the feedstock powder particles (Fig. 1). The quantiles of the volumetric distribution $D_{10}$ = 25 µm, $D_{50}$ = 36 µm, and $D_{90}$ = 50 µm were slightly bigger than those presented by the powder manufacturer (15–45 µm), but still in the optimal spray range for a CS process. The number distribution curve did not deviate much from its volumetric counterpart. This indicated the absence of fine debris particles that typically trigger significant differences between the two curves and cause lower flowability and deposition efficiency.

The spherical shape of the feedstock powder particles with occasional satellites and the dendritic structure shown in Fig. 1 correspond well to the gas atomization process. The grain size appears to be slightly smaller at the outer edges of the particles where the initial solidification occurs during atomization. The grain size obtained from the area fraction shows a nearly Gaussian distribution with a mean grain size around 7 µm (the grain size distribution is presented together with the results on the cold-spray deposits in Fig. 5).

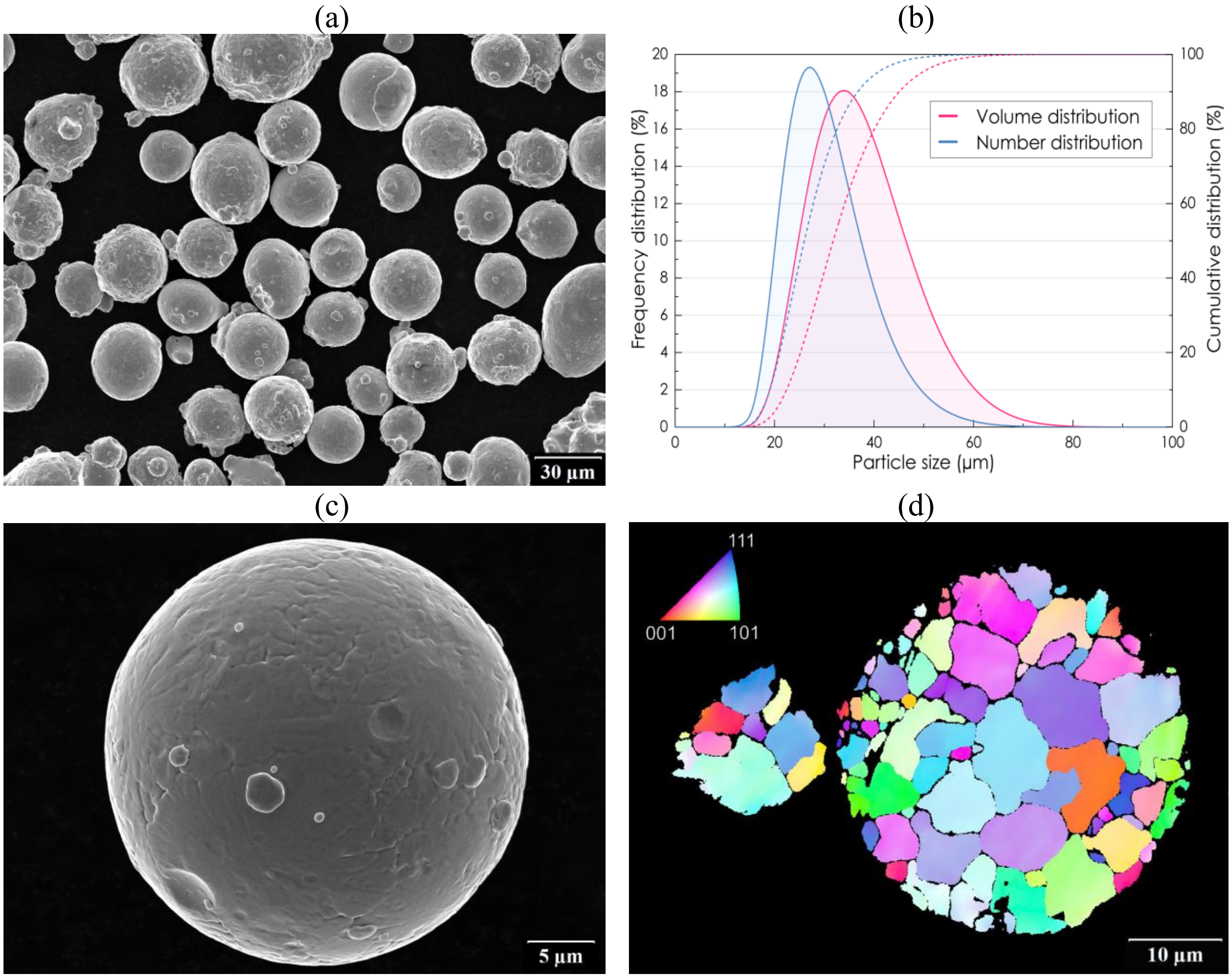

Fig. 1. SEM micrograph of 316L feedstock powders (a), distribution of particle sizes (b), typical powder particle surface morphology (c), and an example of crystallographic orientation in a cross-section (EBSD IPF) (d).

The results of EBSD analysis of cold-spray deposits are shown in Figs. 2–4.

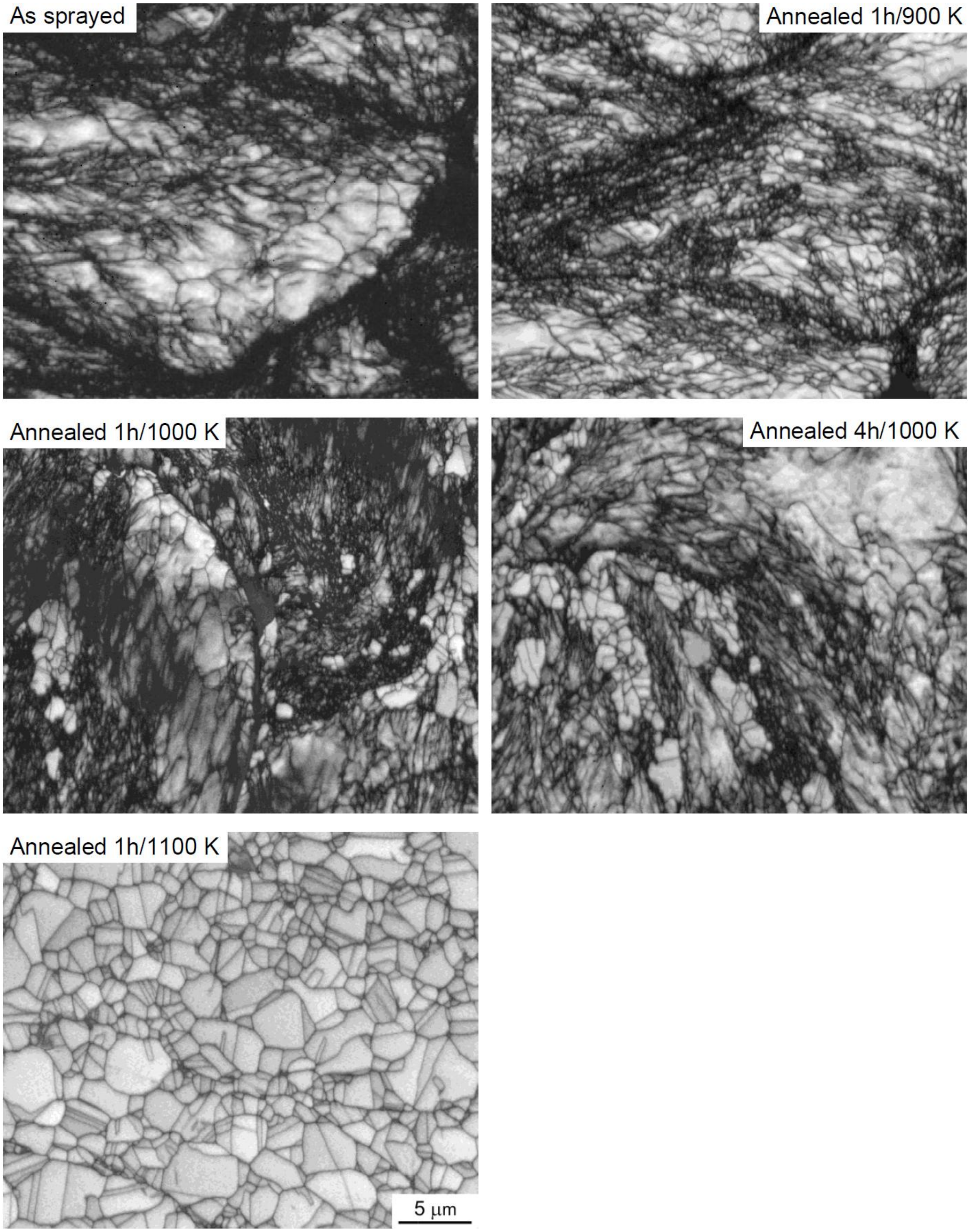

Fig. 2. Grain structure in the cold-sprayed 316L deposits in the as-sprayed and annealed conditions (EBSD CI map).

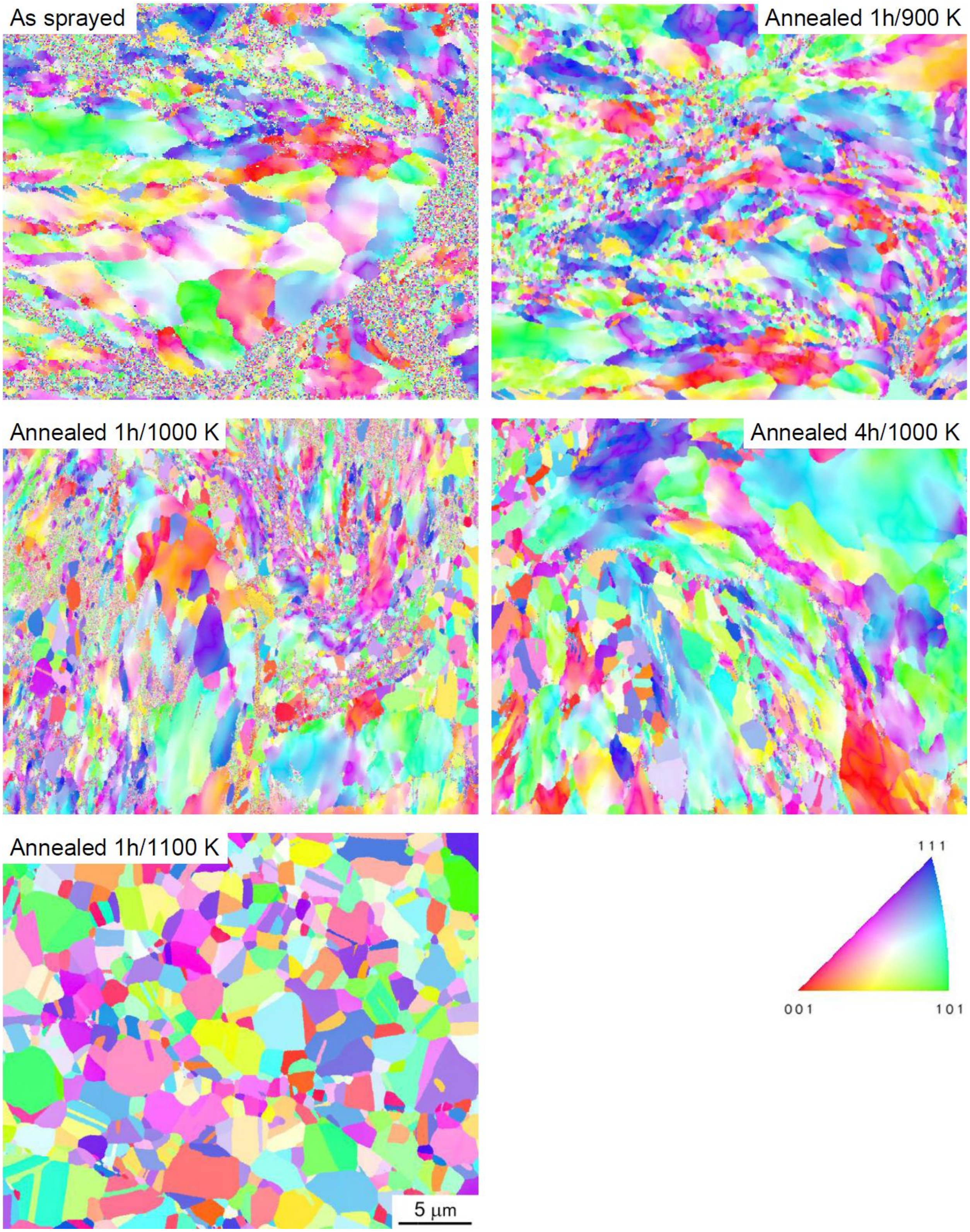

Fig. 3. Crystallographic orientation in the cold-sprayed 316L deposits in the as-sprayed and annealed conditions (EBSD IPF map).

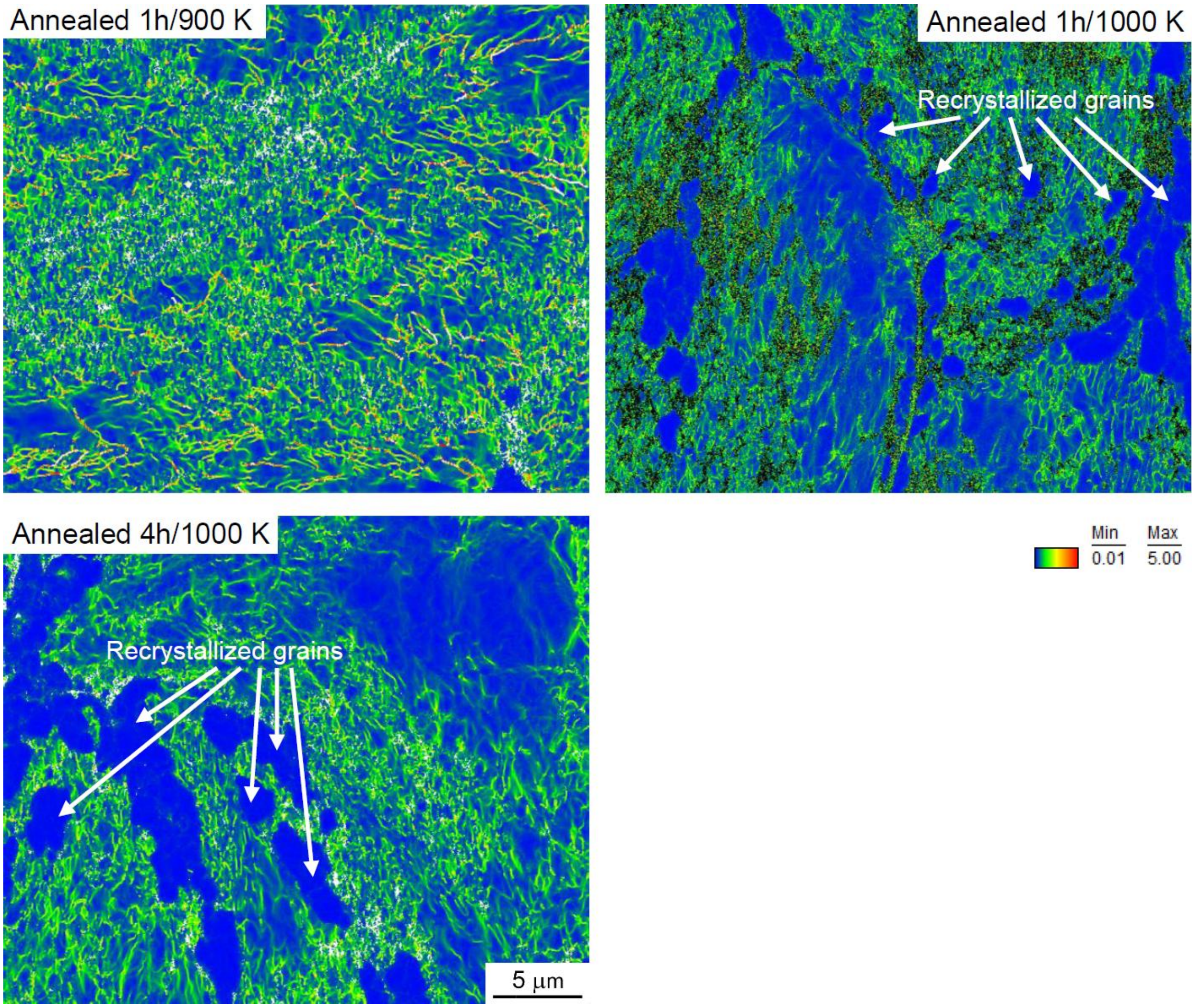

Fig. 4. EBSD KAM map showing distribution of geometrically necessary dislocations in the cold-sprayed 316L deposits after different annealing treatments. The 1 h/900 K treatment corresponds to a state before the onset of recrystallisation. The 1 h/1000 K and 4 h/1000 K treatments illustrate the progressing recrystallisation.

In the as-sprayed condition, the CI maps clearly show a multitude of boundaries and defects created by severe plastic deformation, highly localized at the interfaces between the particles, while the interior of the particles contains many subgrains with low-angle boundaries partly inherited from the original feedstock powder (see IPF maps in Fig. 3). The extremely high density of defects indicates the presence of adiabatic shear instability at the interfaces between the particles, though the EBSD technique itself cannot provide insight into the interior of such fine microstructural features.

The 316L microstructure is practically unchanged after annealing for 1 h at 900 K. At 1000 K, first recrystallized grains appear at the particle boundaries, while in the particle interiors, such recrystallization is not observed and recovery is more likely to occur (compare the KAM maps after annealing for 1 h at 900 K and 1000 K, Fig. 4). The fraction of recrystallized grains increases with annealing time (cf. 5% recrystallized grains after 1 h and 30% after 4 h at 1000 K, Table 2) and temperature; after annealing for 1 h at 1100 K, the microstructure was completely recrystallized.

To obtain statistical data of recrystallized grain sizes ($d_{rex}$), the GOS was used to discriminate the original, deformed grains and the recrystallized grains by setting the average misorientation threshold within the grains as 0.2°. This value, although lower than the thresholds commonly adopted in EBSD recrystallization studies (typically 1–2°), was intentionally chosen to isolate fully strain-free grains and exclude regions that only partially recovered. The selected grains that fulfilled this condition simultaneously exhibited very low KAM values, confirming their recrystallized character. Furthermore, sensitivity tests using higher GOS thresholds (up to 0.5°) produced only minor changes in the resulting grain-size distributions, indicating the robustness of the method.

At 1000 K, the measured grain size corresponds to the early stage of recrystallization, which starts at the particle interfaces (though the EBSD method is unable to detect the very small nuclei, which are therefore not included in the histogram). The recrystallized grain size and fraction gradually increase over time (see Table 2); as the recrystallization spreads inside the particles, the histograms of recrystallized grain size are progressively shifted towards higher grain size values, as can be seen in Fig. 5. At 1100 K, the recrystallization is complete (once all the grains had reached their neighbors and stopped growing after they had replaced the original deformed structure), and the histogram shows the well-developed, heavy-tailed distribution skewed toward larger grain sizes developed inside the original particles. The asymmetric (heavy-tailed) histograms of recrystallized grain size can be well approximated by a log-normal distribution (especially at 1100 K):

$$f(d_{rex}) = \frac{1}{d_{rex}\sigma_d\sqrt{2\pi}}\exp(-\frac{(\ln d_{rex}-\mu_d)^2}{2\sigma_d^2}). \quad (7)$$

The corresponding parameters $\mu_d,\ \sigma_d$ are listed in Table 2:

Table 2. Parameters of the log-normal distribution of the recrystallized grain size after different annealing times and temperatures.

| Annealing temperature /time | $\mu_d$ (μ$m$) | $\sigma_d$ (μ$m$) | Mean (μ$m$) | Modus (μ$m$) | Median (μ$m$) | Recrystallized fraction (%) |
|---|---|---|---|---|---|---|
| 900 K/1 h | - | - | - | - | - | 0 |
| 1000 K/1 h | 0.12 | 0.76 | 1.5 | 0.6 | 1.1 | 5 |
| 1000 K/4 h | 0.34 | 0.61 | 1.7 | 1.0 | 1.4 | 30 |
| 1100 K/1 h | 0.95 | 0.55 | 3.0 | 1.9 | 2.6 | 100 |

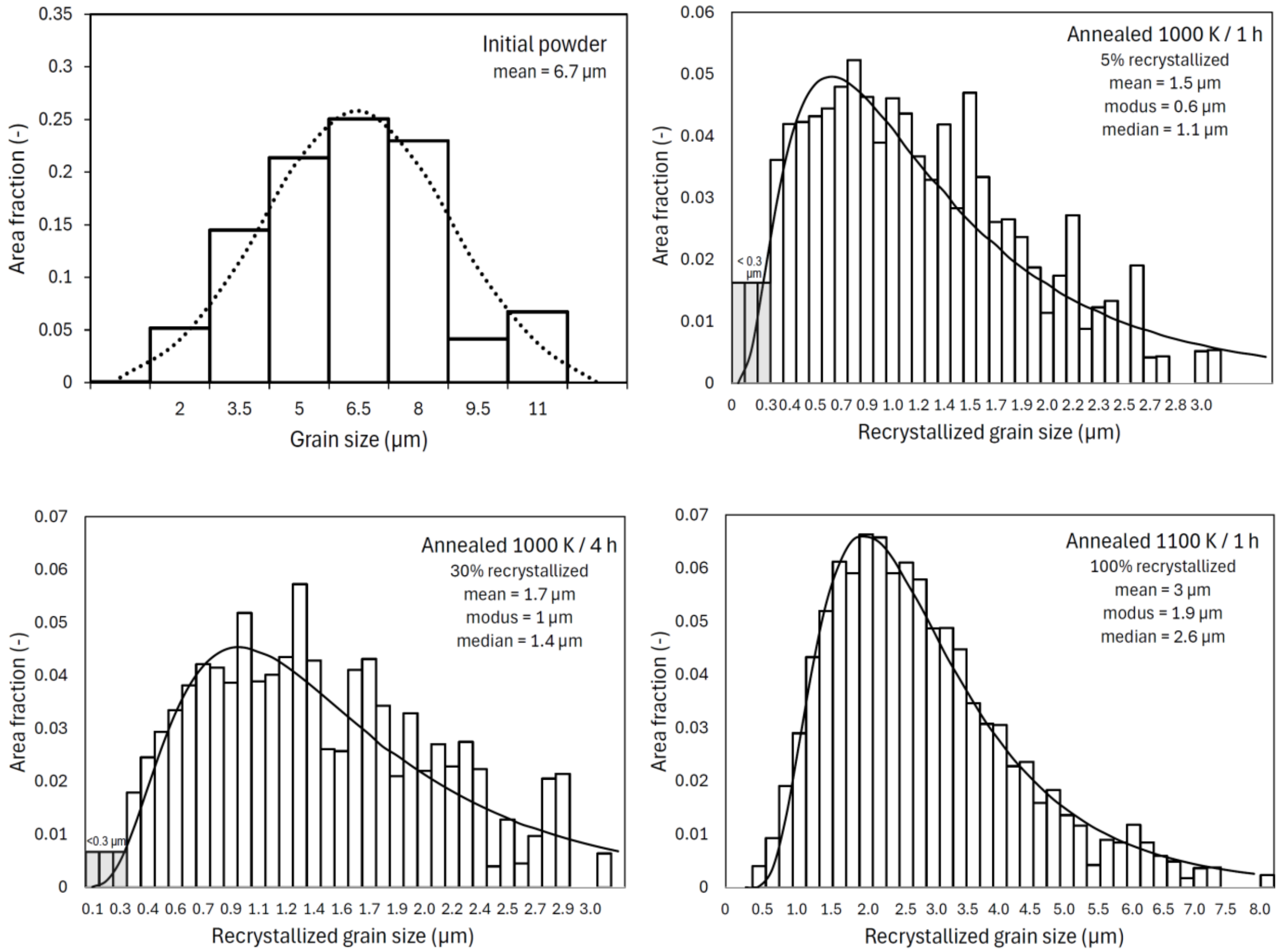

Fig. 5. Histograms of grain size in a representative particle of the feedstock powder and distributions of recrystallized grain size after annealing 1 h at 1000 K, 4 h at 1000 K, and 1 h at 1100 K.

The dependencies of hardness and Young's modulus on depth of penetration obtained using the CMC method are shown in Fig. 6 for different annealing times and temperatures.

The measured Young's modulus remains practically unchanged regardless of the penetration depth, which confirms the reliability of the CMC measurement method. Consistent with the microstructural observations, the dependence of hardness on depth of penetration remains practically unchanged up to the 1000 K/1 h treatment; after that, the hardness begins to decrease as the recrystallization progresses. A clear indentation size effect, i.e., a decrease in hardness with depth, was observed in all samples, although this effect appears to be stronger after recrystallization than in the as-sprayed condition.

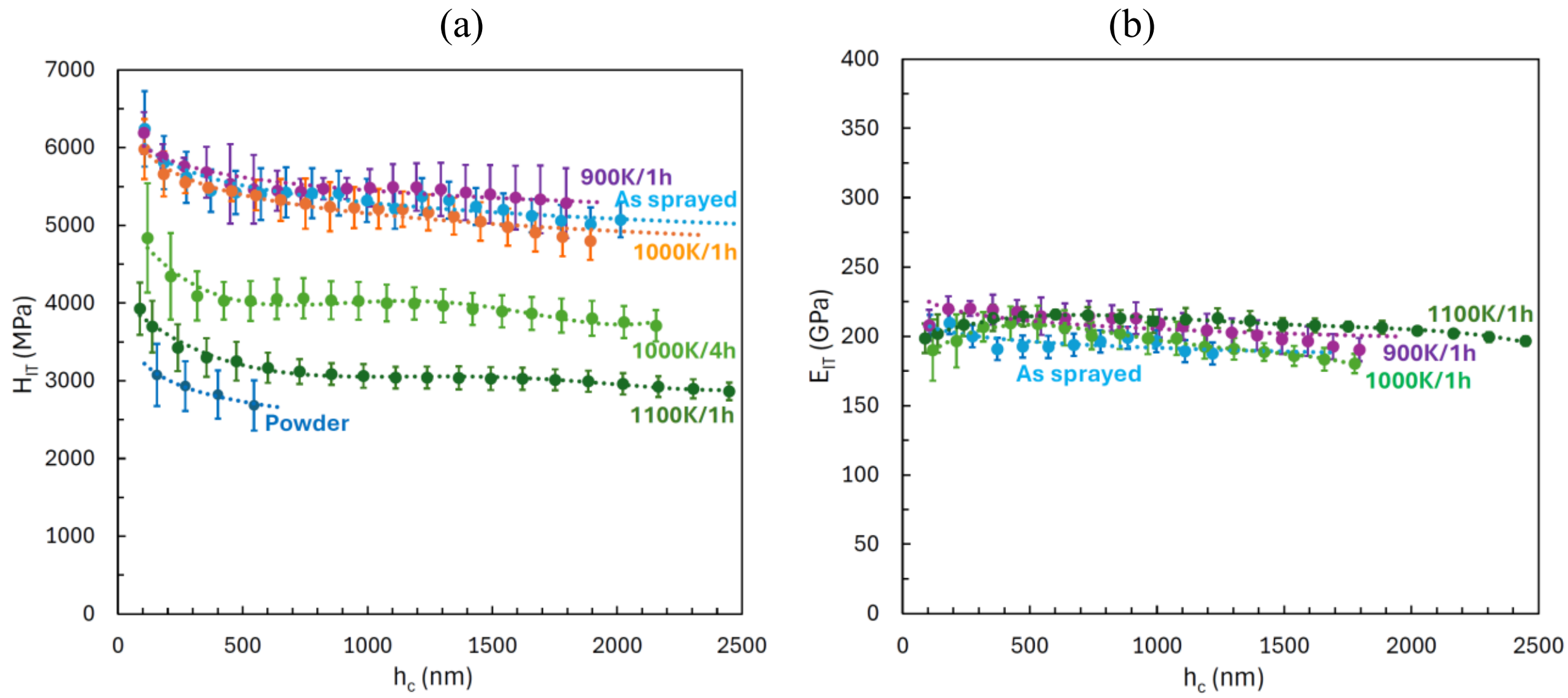

Fig. 6. Hardness (a) and Young's modulus (b) of cold sprayed 316L steel (in different annealing conditions) obtained at different depths of penetration using the CMC method.

The high dislocation density (especially in the as-sprayed condition) tends to suppress the indentation size effect in depths ≥100 nm, as revealed by local measurement by a method without unloading (see Fig. 7a). Complex fine grain/subgrain structure, in turn, can lead to a change in measured hardness as a result of a hard-to-soft (or soft to hard) transition, as the process zone beneath the indenter is "sampling" the properties of multiple grains at once with increasing penetration depth (see Fig. 7b). In the case of heterogeneous materials with structural features of varying sizes and local properties, such as those produced by cold spraying, an apparent size effect may therefore occur purely due to the averaging of hardness at shallow and greater depths.

(b) (b)

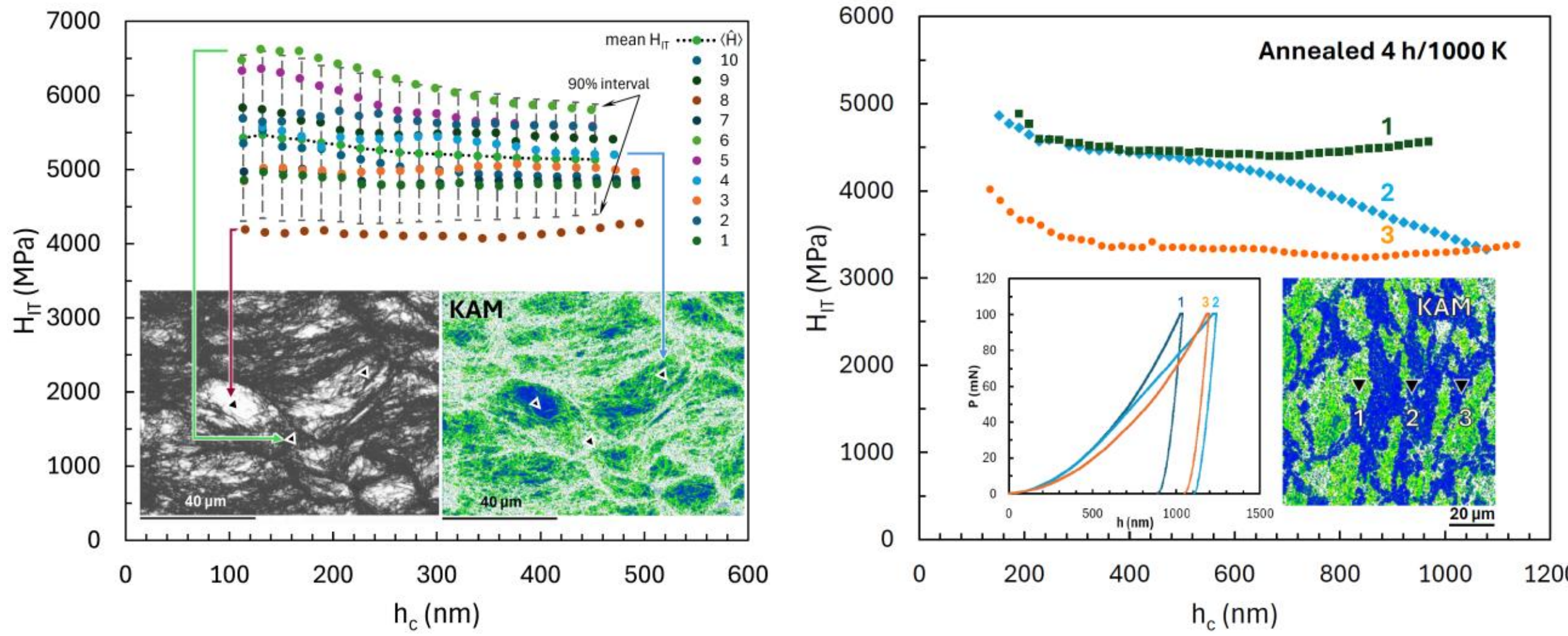

Fig. 7. Local measurements of hardness dependence on depth of penetration obtained using continuous indentation without unloading at locations with different dislocation density in as-sprayed (a) and partially recrystallized 316L steel (b). Selected indentations in areas with different dislocation densities are indicated.

Let's examine the coupling of grain size distribution with spatial partitioning of strain. The experimental data demonstrate that the post-annealed (fully recrystallized) grain size of the 316L stainless steel deposit follows a log-normal distribution with parameters $\mu_d = 0.95$ and $\sigma_d = 0.55$, yielding an arithmetic mean grain size of approximately 3.0 µm. The log-normal distribution is a direct consequence of the severe spatial heterogeneity in plastic deformation and stored dislocation energy induced during the cold spray (CS) deposition. During impact, particles undergo extreme strain localization via adiabatic shear instability. The physical link between the recrystallized grain size ($d_{rex}$) and the prior local equivalent plastic strain in the as-sprayed condition ($\varepsilon$) can be modeled using Sellars' power-law constitutive equations (Ref 26, 27):

$$d_{rex} = A \cdot d_0^a \cdot \varepsilon^{-b}. \tag{8}$$

Using the natural logarithm, we obtain:

$$\ln(d_{rex}) = \ln(A \cdot d_0^a) - b \cdot \ln(\varepsilon), \tag{9}$$

where $d_0$ represents the initial grain size after spraying, $A$ is a material-dependent kinetic constant, and $a$ and $b$ are structural exponents. In the first approximation, the strain exponent $b$ can be taken as 0.5 for austenitic stainless steels, reflecting an inverse square root dependence on dislocation scaling (assuming that Taylor's hardening is predominant). From the relationship between deformation and grain size, it emerges that the deformation will also follow a log-normal distribution. Given the experimentally measured geometric standard deviation of $\sigma_d = 0.55$, the standard deviation of the underlying localized plastic strain distribution ($\sigma_\varepsilon$) is then:

$$\sigma_\varepsilon = \frac{\sigma_d}{b} = 1.10. \tag{10}$$

Such exceptionally high strain variance ($\sigma_\varepsilon = 1.10$) provides clear evidence of the severe microstructural gradients, which are typical of cold sprayed materials. In near-equilibrium thermomechanical processes when thermal and mechanical energies are balanced (e.g., hot rolling), $\sigma_\varepsilon$ rarely exceeds $0.15 - 0.25$. A value of $1.10$ shows that the material underwent severe spatial partitioning of strain, isolating ultra-high strains ($\varepsilon > 3$) at the particle interfaces while leaving the particle cores at lower, near-critical deformation thresholds.

The coupled effects of the inherent extreme strain rates ($\dot{\varepsilon} \approx 10^6 - 10^9\ \mathrm{s}^{-1}$) and the rapid localized heating can be achieved by introducing the temperature-compensated strain rate, the Zener-Hollomon parameter ($Z$):

$$Z = \dot{\varepsilon} \cdot \exp\left(\frac{Q_{def}}{R \cdot T}\right), \tag{11}$$

where $Q_{def}$ is the activation energy for plastic deformation, $R$ is the universal gas constant, and $T$ is the absolute local temperature.

To quantify the adiabatic heating and stored dislocation density at the highly deformed particle interfaces, the transient temperature peak ($\Delta T$) generated by plastic work dissipation can be estimated using the classical Taylor-Quinney equation:

$$\Delta T = \frac{\beta}{\rho \cdot C_p} \int_0^{\varepsilon_{loc}} \sigma \, d\varepsilon = \frac{\beta \cdot \bar{\sigma} \cdot \varepsilon_{loc}}{\rho \cdot C_p}. \tag{12}$$

Using characteristic values for austenitic 316L stainless steel, i.e., a Taylor-Quinney coefficient $\beta = 0.90$, a density $\rho = 8000\,\mathrm{kg/m^3}$, a specific heat capacity $C_p = 500\,\mathrm{J/(kg \cdot K)}$, a high-rate dynamic flow stress $\bar{\sigma} \approx 1000\,\mathrm{MPa}$ (assuming saturation of hardening and the start of the phonon drag regime), and an interfacial local strain $\varepsilon_{loc} = 3$, the calculation yields:

$\Delta T \approx 675\,K$.

Considering that the initial temperatures of the particles at the moment of impact $T_{initial}$ are higher than room temperature but below $475\,\mathrm{K}$ ($\approx 200\,°\mathrm{C}$) (Ref 28), the absolute maximum temperature reached at the particle boundaries during impact ($T_{impact}$) can be estimated as $\sim 1000 - 1150\,K$. Assuming that the strain rates are in the range $\dot{\varepsilon} \approx 10^6 - 10^9\,\mathrm{s^{-1}}$, the activation energy for plastic deformation is approximately $Q_{def} = 400\,\mathrm{kJ/mol}$ for 316L (Ref 29), and the maximum local absolute temperature $T$ is less than 1150 K, the Zener-Hollomon parameter given by Eq. 11 reaches values $Z \approx 10^{24} - 10^{27}\mathrm{s^{-1}}$.

Such extreme values of $Z$ rather suggest that even for very high strains reached at the particle interfaces, the dynamic recrystallization in the 316L materials must have been inhibited during spraying (Ref 30, 31).

The remaining $(1 - \beta)$ = 10% of the total plastic work is stored directly within the lattice as structural defects. This stored energy per unit volume ($E_{stored}$) is about:

$$E_{stored} \approx 0.10 \cdot \int_0^{\varepsilon_{loc}} \sigma \, d\varepsilon = 0.10 \cdot \bar{\sigma} \cdot \varepsilon_{loc} = 3 \times 10^8 \text{ J/m}^3. \tag{13}$$

Equating this bulk energy to the elastic strain energy per unit length of a single dislocation line ($E_{line}$) allows for the direct derivation of the stored dislocation density ($\rho_d$). Using a shear modulus $G = 81\, GPa$, and a Burgers vector $b = 0.254$ nm of 316L:

$$E_{line} = \frac{1}{2} \cdot G \cdot b^2 \approx 2.613 \times 10^{-9} \text{ J/m}, \tag{14}$$

$$\rho_d = \frac{E_{stored}}{E_{line}} \approx 1.15 \times 10^{17} \text{ m}^{-2}. \tag{15}$$

Such a massive dislocation density ($\rho_d \sim 10^{17}$ m$^{-2}$) lies at the upper physical limit achievable by severe plastic deformation (SPD), providing an enormous thermodynamic driving force ($F_{rex} = E_{stored}$) for subsequent microstructural restoration.

According to classical nucleation theory, a newly formed strain-free cell boundary will only grow if the reduction in bulk stored dislocation energy exceeds the energy penalty of creating a new high-angle grain boundary (HAGB). The critical nucleus radius ($R_{crit}$) is defined as:

$$R_{crit} = \frac{2 \cdot \gamma_{HAGB}}{F_{rex}}. \tag{16}$$

Utilizing a characteristic HAGB energy for austenitic steel ($\gamma_{HAGB} = 0.6$ J/m$^2$) (Ref 32), the critical radius and corresponding stable diameter ($D_{crit}$) are $R_{crit} \approx 4$ nm $\Longrightarrow D_{crit} \approx 8$ nm.

Even if nuclei of such size might have been formed in the ASI zones during the spraying, they did not continue to grow as the local temperature dropped rapidly, which would explain why no nuclei could be detected using EBSD, being below the spatial resolution limit.

At the same time, a sub-10 nm critical nucleus size indicates an extreme rate of nucleation at the particle boundaries upon subsequent annealing. This ultra-low critical nucleus size (and corresponding energy barrier) explains why the cold sprayed material achieved complete recrystallization during the 1-hour annealing at 1100 K, whereas conventionally processed 316L undergoes only incomplete recrystallization under these time and temperature conditions. Furthermore, this mechanism explains the geometric dispersion ($\sigma_d = 0.55$) of the log-normal distribution:

- At particle boundaries, the high density of stable sub-10 nm nuclei led to rapid impingement, arresting grain growth early and preserving the ultra-fine grains ($d_{rex} < 1.0\,\mu$m) seen on the left tail of the distribution.
- In particle cores, where lower dislocation densities yielded a larger critical nucleus size and fewer nucleation sites, the grains grew freely throughout the 1-hour duration, forming the larger grain population ($d_{rex} > 7.0\,\mu$m) on the right tail.

An attempt to predict the yield strength ($R_{p0.2}$) can be made via the Hall-Petch relationship (Ref 33, 34):

$$R_{p0.2} = \sigma_0 + k_y \cdot d_{HP}^{-1/2}. \tag{17}$$

Even if utilizing the simple arithmetic mean for grain size ($d_{HP} = 3.0\,\mu m$) can cause an underestimation, as finer grains contribute disproportionately to dislocation pile-ups, using standard parameters for recrystallized austenitic 316L steel (lattice friction stress $\sigma_0 \approx 200$ MPa; Hall-Petch slope $k_y \approx 300\,\text{MPa} \cdot \mu\text{m}^{1/2}$, (Ref 35)) the yield strength can be estimated as $R_{p0.2} \approx 373$ MPa. This yield strength prediction is in very good agreement with the yield stress value that we obtained by compression test of the sample annealed 1 h at 1100 K ($R_{p0.2} = 388$ MPa), although this good match should be treated with caution as the calculated prediction value depends strongly on the selection of Hall-Petch parameters taken from the literature.

The fully recrystallized microstructure of the deposit achieves a yield strength comparable to that of conventional wrought 316L stainless steel (in 1/8 Hard temper), about twice than the minimum guaranteed yield strength of annealed 316L stainless steel (typically 170 MPa). In other words, through the high-velocity cold spray procedure followed by recrystallization annealing at 1100 K, the material can achieve good mechanical strength. As the dislocations generated by cold spray were removed by the recrystallization during annealing, the strength can be solely attributed to the grain refinement.

## Conclusions

In this study, the 316L steel was deposited by cold spray, and the link between microstructural heterogeneity, recrystallization behavior, and mechanical properties was investigated. The obtained results can be summarized as follows:

- Cold-spray deposition of 316L stainless steel produces a highly heterogeneous deformation structure. Severe plastic deformation and defect accumulation are concentrated at particle interfaces, consistent with the occurrence of adiabatic shear instability during particle impact, while particle interiors remain comparatively less deformed.
- Recrystallization behavior directly reflects the spatial distribution of stored strain energy. Recrystallization nucleates first at particle boundaries after annealing at 1000 K, where the highest dislocation density and stored energy are present, and progressively spreads toward particle interiors with increasing annealing time and temperature. Complete recrystallization is achieved after 1 h at 1100 K.
- The recrystallized grain structure follows a log-normal grain-size distribution. The fully recrystallized deposit exhibits a mean grain size of approximately 3 μm and a geometric standard deviation of 0.55, reflecting the strong heterogeneity of deformation introduced during cold spray deposition.

- Analysis of the grain-size distribution provides insight into impact conditions. Using a Sellars-type recrystallization framework, the measured grain-size statistics indicate exceptionally large local strain variations within individual particles, confirming severe strain localization associated with ASI.

- The estimated thermomechanical conditions suggest that dynamic recrystallization is not a dominant restoration mechanism during the cold spraying of 316L steel. Although local temperatures may transiently exceed the conventional recrystallization temperature, the estimated Zener–Hollomon parameters suggest that extensive dynamic recrystallization during particle impact is unlikely. Instead, the stored deformation energy appears to be retained within the microstructure and subsequently released through static recrystallization during post-spray annealing.

- Hardness evolution is governed by recovery and recrystallization. Hardness remains almost unchanged after annealing at 900 K, indicating limited microstructural recovery. Progressive recrystallization at higher temperatures led to a measurable reduction in hardness.

- Controlled recrystallization offers a pathway for property optimization. Despite some softening in comparison to the as-sprayed condition, the fully recrystallized ultrafine-grained microstructure is predicted to achieve a yield strength of approximately 400 MPa through Hall–Petch strengthening.

- Recrystallization can also be used as a practical characterization tool for cold sprayed materials. The evolution of grain size and hardness after annealing provides indirect information about the deformation state generated during particle impact and can be used to assess strain localization in cold-sprayed deposits.

**Acknowledgements**

This work was supported by Ministry of Education, Youth and Sports of the Czech Republic in the frame of the project CZ.02.01.01/00/22_008/0004591 Ferroic Multifunctionalities, co-funded by European Union, and by the Czech Science Foundation project GA26-2310S.

P.H. and J.Č. are members of MecaNano network, supported by COST (European Cooperation in Science and Technology) Action CA21121 and INTERCOST LUC 24135 project (Ministry of Education, Youth and Sports of the Czech Republic).